\pdfoutput=1
\documentclass[%
 reprint,
superscriptaddress,
 amsmath,amssymb,
 aps,
]{revtex4-2}
\usepackage[utf8]{inputenc}
\usepackage[T1]{fontenc}
\usepackage[english]{babel}
\usepackage{verbatim}
\usepackage{microtype} 

\usepackage[normalem]{ulem}
\usepackage{xcolor} 
\usepackage{lipsum} 
\usepackage{dcolumn}
\usepackage{bm}
\usepackage{graphicx} 
    \graphicspath{ {./} } 
\usepackage{textcomp} 
    \usepackage{gensymb} 
\usepackage{siunitx} 
\usepackage{csquotes} 

\usepackage{hyperref} 



\begin{document}

\preprint{APS/123-QED}

\title{Automatic tuning of a donor in a silicon quantum device using machine learning}

\vspace*{0.5cm} 

\author{Brandon Severin}
    \affiliation{Department of Materials, University of Oxford, Parks Road, Oxford, OX1 3PH, United Kingdom}
    \affiliation{School of Electrical Engineering and Telecommunications, University of New South Wales, Sydney, NSW 2052, Australia}
    \affiliation{ARC Centre of Excellence for Quantum Computation and Communication Technology}
\author{Tim Botzem}
    \affiliation{School of Electrical Engineering and Telecommunications, University of New South Wales, Sydney, NSW 2052, Australia}
    \affiliation{Diraq, Sydney, NSW, Australia}
\author{Federico Fedele}
    \affiliation{Department of Engineering Science, University of Oxford, Parks Road, Oxford OX1 3PJ, United Kingdom}
\author{Xi Yu}
    \affiliation{School of Electrical Engineering and Telecommunications, University of New South Wales, Sydney, NSW 2052, Australia}
    \affiliation{ARC Centre of Excellence for Quantum Computation and Communication Technology}
\author{Benjamin Wilhelm}
    \affiliation{School of Electrical Engineering and Telecommunications, University of New South Wales, Sydney, NSW 2052, Australia}
    \affiliation{ARC Centre of Excellence for Quantum Computation and Communication Technology}
\author{Holly G. Stemp}
    \affiliation{School of Electrical Engineering and Telecommunications, University of New South Wales, Sydney, NSW 2052, Australia}
    \affiliation{ARC Centre of Excellence for Quantum Computation and Communication Technology}
\author{Irene Fernández de Fuentes}
    \affiliation{School of Electrical Engineering and Telecommunications, University of New South Wales, Sydney, NSW 2052, Australia}
    \affiliation{ARC Centre of Excellence for Quantum Computation and Communication Technology}
\author{Daniel Schwienbacher}
    \affiliation{School of Electrical Engineering and Telecommunications, University of New South Wales, Sydney, NSW 2052, Australia}
\author{Danielle Holmes}
    \affiliation{School of Electrical Engineering and Telecommunications, University of New South Wales, Sydney, NSW 2052, Australia}
    \affiliation{ARC Centre of Excellence for Quantum Computation and Communication Technology}
\author{Fay E. Hudson}
    \affiliation{School of Electrical Engineering and Telecommunications, University of New South Wales, Sydney, NSW 2052, Australia}
    \affiliation{Diraq, Sydney, NSW, Australia}
\author{Andrew S. Dzurak}
    \affiliation{School of Electrical Engineering and Telecommunications, University of New South Wales, Sydney, NSW 2052, Australia}
    \affiliation{Diraq, Sydney, NSW, Australia}
\author{Alexander M. Jakob}
    \affiliation{ARC Centre of Excellence for Quantum Computation and Communication Technology}
    \affiliation{School of Physics, The University of Melbourne, Parkville, VIC 3010, Australia}
\author{David N. Jamieson}
    \affiliation{ARC Centre of Excellence for Quantum Computation and Communication Technology}
    \affiliation{School of Physics, The University of Melbourne, Parkville, VIC 3010, Australia}
\author{Andrea Morello}
    \affiliation{School of Electrical Engineering and Telecommunications, University of New South Wales, Sydney, NSW 2052, Australia}
    \affiliation{ARC Centre of Excellence for Quantum Computation and Communication Technology}
\author{Natalia Ares}
    \affiliation{Department of Engineering Science, University of Oxford, Parks Road, Oxford OX1 3PJ, United Kingdom}

\date{\today}

\begin{abstract}

    Donor spin qubits in silicon offer one- and two-qubit gates with fidelities beyond 99\%, coherence times exceeding 30 seconds, and compatibility with industrial manufacturing methods. This motivates the development of large-scale quantum processors using this platform, and the ability to automatically tune and operate such complex devices. In this work, we present the first machine learning algorithm with the ability to automatically locate the charge transitions of an ion-implanted donor in a silicon device, tune single-shot charge readout, and identify the gate voltage parameters where tunnelling rates in and out the donor site are the same. The entire tuning pipeline is completed on the order of minutes. Our results enable both automatic characterisation and tuning of a donor in silicon devices faster than human experts.
\end{abstract}

\maketitle

\section*{Introduction}

    The promised treasures held within silicon ion-implanted donor spin qubits have led to their repeated pursuit as a contending scalable, CMOS-compatible quantum computing architecture ever since their proposition in 1998 \cite{Kane1998}. Exploiting silicon’s central role in the electronics industry and its mature fabrication techniques, placing a $^{31}$P ion in natural~\cite{Jamieson2005, Jakob2022y, Jakob2024x} and isotopically enriched $^{28}$Si material \cite{Lim2025z}, has enabled milestones such as single shot electron spin readout \cite{Morello2010}, 30-second coherence times \cite{Muhonen2014} and error correction threshold fidelities \cite{Madzik2022}. However, the realisation of quantum computers built upon ion-implanted donors in silicon \cite{Jakob2022y, Jakob2024x} is not only ‘dependent on future refinements of conventional silicon electronics’ \cite{Kane1998}, but also the development of automatic approaches to tune, optimise and control such devices in CMOS compatible architectures~\cite{AIreview, Steinacker2025}.
 
    In this work, we take the first step to address this challenging task by introducing \texttt{donorsearch}, an algorithm that automatically tunes an ion-implanted donor device from a de-energised device to a regime poised for spin-selective readout. To achieve this, our algorithm automatically sets up a single-electron transistor (SET) for charge sensing readout, locates the donor's charge transitions, tunes single-shot charge readout, and optimises the rates at which electrons tunnel on and off the donor site to be nearly equal to each other within the readout bandwidth. This establishes a reference operating regime for spin qubit operation.
    
    Until now, automatic methods have only catered to tuning gate-defined quantum dot devices \cite{Baart2016, Botzem2018, Mills2019, Moon2020, VanEsbroeck2020, Darulova2020, VanStraaten2022, Severin2021, hickie2023, Schuff2022, Schuff2024}. Our algorithm utilises computer vision and embedded unsupervised machine learning-enabled methods to process and classify quantum transport signals synonymous with donors in silicon devices.

    Unlike in gate-defined quantum dots, where gate voltages gradually modify the confinement potential and lead to smooth, gradual changes in the transport features observed in charge stability diagrams, tuning an ion-implanted donor device is a true needle-in-a-haystack problem. The extensive gate voltage space has to be searched to find a single charge transition, associated with electrons tunnelling on and off a single level associated with the implanted donor ion. Then, gate voltages need to be finely tuned to achieve single-shot readout of charge and spin. To tackle this tenacious tuning task, \texttt{donorsearch} comprises three stages: a "coarse tuning" stage to tune the SET and acquire the donor-SET charge stability diagrams, a "Handshake" to locate the charge transitions and find readout thresholds, and a "fine tuning" stage to tune single-shot charge readout. Machine learning search and classification methods are leveraged in the fine tuning stage to efficiently navigate the parameter space and tune tunnel rates. Importantly, the algorithm exhibits accuracies in excess of 77\% compared to the unanimous vote of human labels. Tuning times are as short as 10 minutes, notably faster than human experts. Our work thus opens the door to a robust, general, and scalable approach for tuning and optimising donor spin-qubit devices.

\section*{Methods}
    
    \subsection*{The device}
    
        \begin{figure}[t]

            \includegraphics[width=0.99\columnwidth]{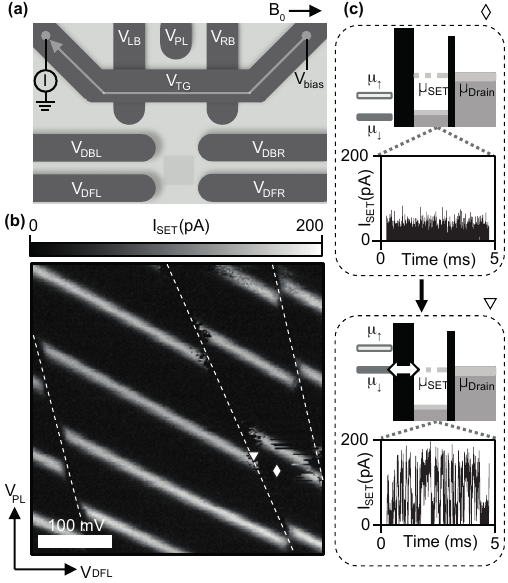}
             
            \caption{(a) Ion-implanted donor in a silicon device.
            Charge sensing readout is carried out by measuring the current across an SET defined by gate voltages $V_{\mathrm{TG}}$, $V_{\mathrm{LB}}$, $V_{\mathrm{RB}}$, and $V_{\mathrm{PL}}$, with $V_{\mathrm{bias}}$ the source-drain bias. 
            Four gate voltages, $V_{\mathrm{DFL}}$, $V_{\mathrm{DFR}}$, $V_{\mathrm{DBL}}$, and $V_{\mathrm{DBR}}$, control the electrostatic potential around the ion-implantation site window center, indicated with a dark grey square. (b) Charge stability diagram acquired by measuring the SET current $I_\mathrm{SET}$ while sweeping $V_{\mathrm{PL}}$ and $V_{\mathrm{DFL}}$. White dashed lines mark charge transitions between the SET and several donor sites, as suggested by the different line slopes.
            Within a charge stability diagram, each donor-SET transition is visited by the algorithm. Here, the SET current is measured at different locations (e.g. at the triangular and diamond markers), to build a telegraph signal classifier that would later be used during the fine tuning stage. 
            (c) During the fine tuning stage, gate voltages are adjusted to achieve an approximately equal proportion of tunnelling-in and tunnelling-out events. This corresponds to the regime where the electron spin-down electrochemical potential ($\mu_{\downarrow}$) aligns with that of the SET ($\mu_{\mathrm{SET}}$), allowing for efficient loading and unloading of spin-down electrons in the presence of a magnetic field B$_\mathrm{0}$, of 1.1 T.}    
            \label{fig:fig1_devices}
        \end{figure}

        We demonstrate \texttt{donorsearch} on a $^{31}$P/$^{123}$Sb co-ion-implanted device in silicon with a proximal single electron transistor (SET) for readout \cite{Morello2010} (Fig.~\ref{fig:fig1_devices}a).  The $^{31}$P donor is shallowly placed ($\sim$7 nm) beneath the gate oxide interface via 8 keV P ions \cite{Jakob2022y, Holmes2024a} within an electron-lithographic implantation window, which is centered on a region outlined in Fig.~\ref{fig:fig1_devices}a. Similarly, the $^{123}$Sb donor is placed ($\sim$5 nm) beneath the gate oxide interface via 10 keV Sb ions \cite{Jakob2024x}. The electrochemical potential of the electron bound to the implanted donor is controlled by electrostatic voltages applied to the donor gate electrodes, $V_{\mathrm{DFL}}$, $V_{\mathrm{DFR}}$, $V_{\mathrm{DBL}}$ and $V_{\mathrm{DBR}}$. At the process level, the stack (ion implantation into enriched $^{28}$Si, SiO$_2$ dielectric, Al top gates, rapid thermal anneal) follows modules that are compatible with CMOS pilot-line fabrication.

        During manual tuning, a human would typically acquire a donor-SET charge stability diagram (Fig. \ref{fig:fig1_devices}b).
        These diagrams are obtained by sweeping a donor gate voltage ($V_{\mathrm{DBL}}$) and the SET plunger gate voltage ($V_{\mathrm{PL}}$), while measuring $I_\mathrm{SET}$. 
        Peaks in current correspond to the SET Coulomb peaks. The relative capacitive coupling of $V_{\mathrm{DBL}}$ and $V_{\mathrm{PL}}$ to the SET can be extracted from the gradient of the Coulomb peak slopes. Breaks in the Coulomb peaks indicate a donor-SET charge transition, i.e., a charge transition where an electron tunnels between the donor site and the SET. In Fig.~\ref{fig:fig1_devices}b, we attribute different sets of transitions, identified with white dashed lines, to different donors within the implantation area.
        
        In the presence of a magnetic field (B$_\mathrm{0}$), the spin degeneracy is lifted, resulting in a splitting of the spin-up ($\mu_{\uparrow}$) and spin-down ($\mu_{\downarrow}$) electrochemical potentials by the Zeeman energy (Fig. \ref{fig:fig1_devices}c).
        To achieve single-shot charge readout, which is a necessary requirement to enable spin readout via Elzerman spin-to-charge conversion~\cite{Elzerman2004}, a human would adjust the gate voltages until electron tunnelling events are observed in the vicinity of a donor-SET charge transition. This indicates that the observed charge dynamics is slow enough relative to the measurement bandwidth to allow the detection of individual tunnelling events. In this regime, the alignment of the spin-down electrochemical potential $\mu_{\downarrow}$ with that of the SET ($\mu_{\mathrm{SET}}$) gives rise to a telegraph signal in the SET current $I_\mathrm{SET}$~\cite{Morello2009, Morello2010}, a hallmark of single shot charge readout (see Fig. \ref{fig:fig1_devices}c). 
        
        \texttt{donorsearch} identifies the donor-SET charge transitions within the charge stability diagram, tunes the gate voltages to enable single-shot charge readout, then fine tunes to achieve $\mu_{\downarrow} = \mu_{\mathrm{SET}}$ (Fig. \ref{fig:fig1_devices}c). When this condition is achieved, the tunnel rates to load and unload a spin-down electron to the donor site are nearly equal, thus providing a fiducial point for enabling efficient spin-readout and qubit operations. All the experimental runs were carried out in a dilution refrigerator at a base temperature of 20 mK and a B$_\mathrm{0}$ of 1.1 T oriented along the SET accumulation channel (see Fig. \ref{fig:fig1_devices}a).

    \subsection*{The \texttt{donorsearch} algorithm} 

        \begin{figure*}
        \includegraphics[width=\textwidth]{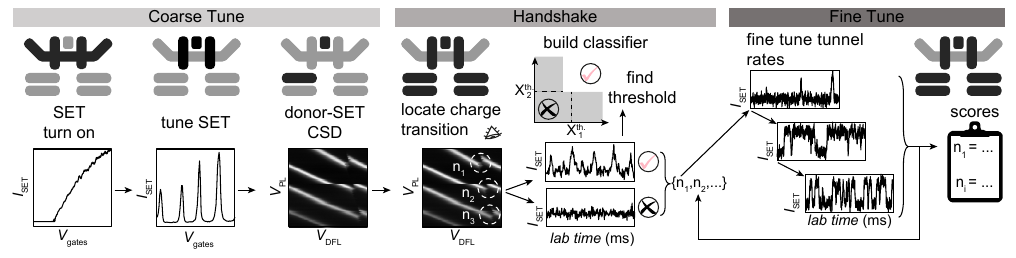}
        \caption{\label{fig:fig2_algo_workflow}
        Outline of \texttt{donorsearch}'s workflow. The coarse tuning stage characterises, checks the functioning of, and tunes the SET,  as well as acquiring a charge stability diagram of the donor-SET gate voltage space. The Handshake stage consists of signal processing routines to locate donor-SET charge transitions within the acquired stability diagram. Multiple current traces are acquired in the vicinity of the donor-SET charge transitions to build a classifier based on two thresholds $X^{th}_{\mathrm{1}}$ and $X^{th}_{\mathrm{2}}$, utilised in the fine tuning stage. In the fine tuning stage, the algorithm takes control of 7 out of the 8 gate voltages and attempts to tune each donor-SET charge transition ($n_{\mathrm{i}}$) to the point of observing a telegraphic signal in $I_\mathrm{{SET}}$. $V_{\mathrm{bias}}$ is fixed before commencing the algorithm.}
        \end{figure*}
        
        The \texttt{donorsearch} algorithm consists of three main stages: coarse tuning, a Handshake, and fine tuning (see Fig. \ref{fig:fig2_algo_workflow} and the Supplementary material for additional information). Each stage progressively focuses on a reduced number of regions in the voltage space that may realise a donor electron spin qubit. Before starting \texttt{donorsearch}, $V_{\mathrm{bias}}$ is manually set to 1 mV and is fixed throughout the entire tuning procedure. This value for $V_{\mathrm{bias}}$ was chosen due to our prior experience in tuning other devices of the same architecture.  All gate voltages are initialised to a value of 0 V.
        
        During the coarse tuning stage, the algorithm first checks for the turn-on of the SET by measuring the SET current, $I_{\mathrm{SET}}$, as a function of $V_{\mathrm{LB}}$, $V_{\mathrm{RB}}$ and $V_{\mathrm{TG}}$, then proceeds to tune the SET gates ($V_{\mathrm{LB}}$, $V_{\mathrm{RB}}$, $V_{\mathrm{PL}}$) while keeping $V_{\mathrm{TG}}$ fixed at 2 V, until the Coulomb blockade regime is reached. It then explores the donor-SET charge stability diagram by measuring $I_{\mathrm{SET}}$ while sweeping $V_{\mathrm{PL}}$ and $V_{\mathrm{DFL}}$, with the remaining donor gates ($V_{\mathrm{DBL}}$, $V_{\mathrm{DBR}}$, and $V_{\mathrm{DFR}}$) held at 0 V. A more exploratory mode of this stage of the algorithm can also be used, in which charge stability diagrams are acquired for other values of $V_{\mathrm{DBL}}$, $V_{\mathrm{DBR}}$, and $V_{\mathrm{DFR}}$. Each resulting stability diagram is passed to the Handshake stage, the next step of the tuning algorithm.

        The Handshake stage comprises a charge transition locator module and the telegraphic signal classifier (see Supplementary material for further information). The charge transition locator module takes donor-SET charge stability diagrams as inputs and uses image analysis and computer vision techniques to locate donor-SET charge transitions. At each identified transition, we select a point at the SET-Coulomb peak discontinuity (see Supplementary material). Starting from the gate voltage configuration at this location, the algorithm then generates ten random configurations by independently adjusting each gate voltage by an amount uniformly sampled from a $\pm$ 2~mV range. This restricted search space increases the likelihood of observing current traces with tunnelling events, enabling efficient training of the telegraphic signal classifier.
        For each of these ten configurations, the algorithm acquires a 30 ms current trace by recording $I_{\mathrm{SET}}$ at a sampling rate of 500 kHz, while holding all gate voltages constant during acquisition. These traces are then classified using a K-means clustering algorithm with a target of two clusters. One with traces exhibiting telegraphic signals and one with traces without such a feature. The approximate boundaries between the two clusters $X^{th}_{\mathrm{1}}$ and $X^{th}_{\mathrm{2}}$ are then used as a fast classification during the fine tuning stage. 

        In the fine tuning stage, the algorithm takes control of all gate voltages except $V_{\mathrm{TG}}$, which remains fixed at 2 V. The algorithm explores the gate voltage space by independently adjusting each gate voltage value within a range of $\pm$ 5 mV from the identified location of the donor-SET charge transitions, while acquiring 30 ms current traces at each point, with the aim of detecting and fine tuning the telegraphic signals. At this stage, to search the gate voltage space, the algorithm uses Gaussian process Bayesian optimisation, similar to the approach taken in \textit{Harmless Bayesian Optimisation} \cite{Ahmed2016}. To benchmark our results, we use a simple random search, since it is straightforward to implement, relies on no heuristics, and makes no assumptions about the underlying search space. 
        
        Each current trace acquired at this stage is first analysed by the telegraphic signal classifier. The worst algorithm score (Sc) is 1, which is assigned by the telegraphic signal classifier if no tunnelling events are found. If tunnelling events are detected, a dynamic threshold is extracted from a double Gaussian fit applied to the trace histogram. The algorithm then estimates the amount of time spent above ($T_\mathrm{a}$) and below ($T_\mathrm{b}$) this threshold, and generates the trace score using the normalised absolute difference between $T_\mathrm{a}$ and $T_\mathrm{b}$, relative to the 30 ms trace duration (see Supplementary material). The goal of the Gaussian process Bayesian optimisation search is then to locate the coordinates in the gate voltage space with the lowest score value, corresponding to a score of 0, where $T_\mathrm{{a}} = T_{\mathrm{b}}$.
        
        The algorithm requires a score lower than 0.1 to stop the search and proceed to fine tune the next donor-SET charge transition. This means that $T_\mathrm{{a}}$ and $T_{\mathrm{b}}$ differ by no more than 10\% of the total current trace duration, corresponding to a difference of 3 ms. The algorithm samples up to 50 points in gate voltage space at each transition site before attempting to fine tune the next transition present in the charge stability diagram. The algorithm continues to explore and attempt to tune each transition present in the charge stability diagram one by one. The donor-SET charge transition coordinates that were successfully fine tuned to optimal electron tunnel rates are output by the algorithms. These can then be visited by the user for spin-readout calibration~\cite{Morello2010, Morello2009}. 
        
    \section*{Results}
    \begin{figure}[t] 
        \includegraphics[width=\columnwidth]{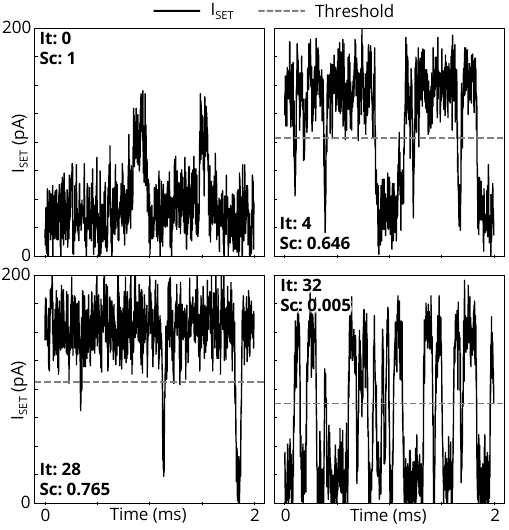}
        \caption{\label{fig:fig5_search_for_rts} Current traces acquired during the search for telegraphic signals, showing the first 2 ms of four current traces acquired at various iterations of the fine tuning stage (It: 0, 4, 28, 32) at one particular donor-SET charge transition. Each current trace is scored (Sc). Current traces that do not exhibit any clear tunnelling events or have low current ($\sim 100$ pA) are classified as noise (for example, iteration 0) and receive a score of Sc = 1. 
        Those that are classified as telegraphic signals are scored using a dynamic current threshold (dashed line). The condition for optimal charge readout is achieved once a score Sc $<$ 0.1 is reached.}
    \end{figure}

    The algorithm was run 28 times from a completely de-energised device, with one thermocycle after 14 runs. Of the 28 runs, 14 were conducted using the Gaussian process Bayesian optimisation during the fine tuning stage of the algorithm, while the remaining 14, used as a benchmark, were performed using random search instead. In Fig. \ref{fig:fig5_search_for_rts}, we show current traces acquired for a particular donor-SET charge transition during the fine tuning stage using Gaussian process Bayesian optimisation. Traces deemed by \texttt{donorsearch} to exhibit telegraphic signals were labelled by humans to gauge the algorithm's performance.  

    On average, 13 donor-SET charge transitions were detected in each run, with a standard deviation of 2 transitions across the 28 experimental runs. To separately gauge the performance of the telegraphic signal classifier, two humans labelled the current traces collected from the Handshake stage during one of \texttt{donorsearch} experimental runs. 
    The classifier achieved a balanced accuracy score of 88\% when compared to the unanimous vote of human labels for a current trace deemed to show single-shot readout.
    This corresponds to the point (0.03, 0.80) on the receiver operating characteristic curve (ROC). A balanced accuracy of 86\% was achieved when compared to the unanimous vote of human labels for traces that were classified as noise, corresponding to a ROC point of (0.28, 0.99). 
     
    In the fine tuning stage, over the 28 runs, we found the average value of the dynamic threshold to be $73 \pm 9.5$ pA when considering all the traces that were unanimously found to exhibit single-shot readout by the human labellers. When comparing the unanimous human labels to the current traces output by \texttt{donorsearch}, it achieves an accuracy of 77\%. 
    
    As we show in Table~\ref{tab:accuracy}, if we compare the performance of the algorithm against the benchmark runs, the Gaussian process Bayesian optimisation results in an accuracy of 77\% against the random search with an accuracy of 50\%. 
    If we remove the unanimous vote requirement of the human labels and rely on a single human label confirming the presence of single-shot readout, these accuracy scores increase respectively to 90\% and 70\%. This discrepancy has to do with the different judgement of the labellers.

    \begin{table}[t]
    \begin{ruledtabular}
    \begin{tabular}{cccc}
    \multicolumn{1}{l}{} & \multicolumn{2}{c}{\texttt{donorsearch} accuracy} \\
    Labellers Agreement & Gaussian Process  & Random Search \\ \hline
    One Vote  & 90\%   & 70\%  \\
    Unanimous  & 77\%   & 50\%  \\
    \end{tabular}
    \end{ruledtabular}
    
    \caption{\label{tab:accuracy} \texttt{donorsearch} accuracy: Donor search accuracy broken down by human labellers' agreement. One vote corresponds to one of the two labellers labelling a current trace output as a positive (i.e. the trace exhibits random tunnelling events). Unanimous corresponds to both labellers labelling a current trace output as a positive. }
    \end{table}

    \begin{table}[t]
    \begin{ruledtabular}
    \begin{tabular}{ccc}
    \multicolumn{1}{l}{} & \multicolumn{2}{c}{Tuning Times (s)} \\
    Stage                & Gaussian Process  & Random Search    \\ \hline
    Coarse          & \multicolumn{2}{c}{318 (315, 323)}  \\
    Handshake            & \multicolumn{2}{c}{117 (104, 145)}  \\
    Fine            & 132 (108, 163)   & 228 (167, 329)  \\ \hline
    Total Median              & 567   & 663
    \end{tabular}
    \end{ruledtabular}
    \caption{\label{tab:tuning_times} Median device tuning times broken down by each stage of the algorithm with 80\% credible intervals (equal-tailed). \texttt{donorsearch} fine tuning stage tuning times are split based on the search method used, Gaussian process Bayesian optimisation or random search.}
    \end{table}

    In table~\ref{tab:tuning_times}, we present the tuning times for the fine tuning stage calculated using the same Bayesian multilabeller statistics used in Ref. \cite{Moon2020}. Only the fine tuning stage was performed using two different methods, and that is why we report a value for each method in this case. The median tuning times for the coarse and Handshake stages are 318 (315, 323) and 117 (104, 145) seconds with 80\% credible intervals (equal-tailed), respectively.
    The majority of the time taken by the coarse tuning stage is dedicated to the SET tuning, as the acquisition of the donor-SET charge stability diagram takes only 25 seconds out of the total of 318 seconds.
    The Handshake stage is the fastest part of the tuning algorithm, taking a median time of 117 seconds. The charge transition locator module takes 0.61 seconds and the remaining time is used to build the noise classifier. Building the classifier takes the longest time in this stage. Different build times depend on the number of charge transitions found in the previous step.
    The fine tuning stage is the second fastest part of the \texttt{donorsearch} algorithm, taking a median time of 132 seconds. The machine learning-based search method is 40\% faster than the random search benchmark runs when comparing the median tuning times. 

\section*{Discussion}
    In the Handshake stage, we attribute the variations in the number of transitions identified by the charge locator module across different runs primarily to the harsh filtering required to reduce the number of false positive transitions (see Supplementary material for further details). The module also extracts the values of the slopes of the SET-Coulomb peaks, which provide useful information for calibrating spin-readout and qubit control operations~\cite{Morello2009, Morello2010}. 
    The telegraph signal classifier has the benefit of being both accurate and built in situ with a minimal number of training samples. A region of gate voltage space may contain donor-SET charge transitions, but they may not be tuned to a point where single-shot charge readout can be detected. 
    The time required to build the classifier could be reduced by acquiring fewer current traces near each transition. 
    It is, however, unlikely that fewer traces would provide a sufficiently diverse set for the K-means clustering algorithm; hence, this approach would most likely require a different classification method. 
    Alternative classification methods were explored, including comparisons between single- and double-Gaussian fit quality, and the use of a fixed current threshold. However, these approaches lacked robustness, either failing to detect telegraphic signals with small amplitudes or allowing too many false positives.
    Nonetheless, our unsupervised embedded learning method is able to create a classifier that has an in-the-field accuracy of 86\% in under two minutes, without the inclusion of augmented or simulated data and while being agnostic to the overall device architecture. 
    
    In the identification of single-shot readout, experimental runs using our Gaussian process Bayesian optimisation method significantly outperformed in accuracy the benchmark runs based on random search by at least 20\%, regardless of the voting metric chosen by the human labellers.
    The observation that Gaussian process Bayesian optimisation outpaces random search in tuning speed is consistent with previous studies on machine learning-assisted coarse tuning of gate-defined double quantum dots in GaAs \cite{Moon2020}, as well as in silicon- and germanium-based architectures \cite{Severin2021, VanStraaten2022}.
    Although the limits of the fine tuning search space are confined to 10~mV around the located transitions, the high dimensionality of the search space (7 gate voltages), prevents the random search from outperforming a more efficient search method. Even though the fastest random search tuning times venture into the range of that achieved by Gaussian process Bayesian optimisation, these occurrences are rare given that the 80\% confidence interval of random search benchmark runs (162 seconds) is three times wider than that of our method based on Gaussian process. These results show that our machine learning-enabled search methods result in both reliable and fast tuning of quantum devices. 

\section*{Conclusion}
    \texttt{donorsearch} is the first algorithm of its kind capable of tuning single-shot charge readout in an ion-implanted donor in a silicon device from scratch~\cite{AIreview}. Tuning times are on the order of minutes, putting \texttt{donorsearch} ahead of human experts tuning such devices. 
    \texttt{donorsearch} incorporates both coarse and fine tuning stages into a single package, including a donor-SET charge transition locator module that does not require any pretraining. The fine tuning stage is aided by an efficient and lightweight telegraphic signal classifier, which is built using unsupervised embedded learning, enabling fast tuning times. 

    Similar methods could be used to automate stages such as spin-readout calibration~\cite{Schuff2022, Carlsson2025}, relaxation-time measurements~\cite{Ansaloni2023} and electron spin resonance~\cite{Schuff2024}. Moreover, an automated and efficient approach to benchmarking and optimising properties such as Rabi frequencies and spin readout could help clarify the currently poorly understood challenges associated with tuning these parameters \cite{Johnson2021, Carlsson2025}.
    By pairing CMOS-compatible fabrication with control-layer automation, the previously unexplored waters of automatic tuning in ion-implanted donors in silicon, soon to be densely charted, will open doors to scalability and a deeper understanding of these devices.

\section*{Data Availability}
    The data acquired by the algorithm during experiments is available online from the following URL: \url{https://doi.org/10.5281/zenodo.17451553} 
    
\begin{acknowledgments}
 This work was supported by the Royal Society, the EPSRC National Quantum Technology Hub in Networked Quantum Information Technology (EP/M013243/1), Quantum Technology Capital (EP/N014995/1), EPSRC Platform Grant (EP/R029229/1), the European Research Council (Grant agreement 948932). Work at UNSW was funded by the Australian Research Council (grant no. CE170100012). A.M. acknowledges support of an Australian Research Council Laureate Fellowship (grant no. FL240100181). X.Y., B.W. and H.G.S. acknowledge support from the Sydney Quantum Academy. Devices were fabricated using the facilities at the UNSW node of the Australian National Fabrication Facility (ANFF).
\end{acknowledgments}

\section*{Author Contributions}

    B.S. and the machine performed the experiments at the University of New South Wales. T.B., X.Y., B.W., H.G.S., and D.S. contributed to the experiment. I.F.d.F., D.H., A.M.J., F.E.H., A.S.D., and D.N.J. contributed to the sample fabrication. B.S. developed the algorithm. T.B. contributed to labelling and data analysis. The project was conceived by B.S., A.M. and N.A.. B.S., F.F. and N.A. wrote the manuscript. All authors commented and discussed the results.
    
\section*{Competing Interests}
    Brandon Severin declares a competing interest as a founder of Conductor Quantum, Inc., which develops machine learning-based software for semiconductor quantum devices. Natalia Ares declares a competing interest as a founder of QuantrolOx, which develops machine learning-based software for quantum control. All remaining authors declare no conflicts of interest. 
    
\section*{Correspondence}
    Correspondence and requests for materials
    should be addressed to Natalia Ares \\  
    (email: \href{mailto:natalia.ares@eng.ox.ac.uk}{natalia.ares@eng.ox.ac.uk}).


\bibliographystyle{naturemag}
\bibliography{ms.bib}

\clearpage

\section*{Supplementary Material}
\setcounter{equation}{0}
\setcounter{figure}{0}    
\setcounter{table}{0}
\renewcommand{\figurename}{Supplementary Figure}
\renewcommand{\tablename}{Supplementary Table}

\renewcommand\tablename{Supplementary Table}
\renewcommand\figurename{Supplementary Figure}

    \subsection*{Supplementary Methods}
        
        \subsubsection*{Experimental setup and control}\label{sup:setup}
        The device consisted of a natural silicon wafer topped with a 900 nm epitaxial layer of isotopically enriched $^{28}$Si. A two-step thermal oxide SiO$_2$, separated the $^{28}$ Si-enriched substrate and Al gate electrodes on the top surface of the device, which were patterned using electron beam lithography. The implantation of the $^{31}$P donor was performed via 8 keV P ions (and $4\times10^{11}\ \mathrm{cm^{-2}}$ fluence), and the implantation of the $^{123}$Sb donor was performed via 10 keV Sb ions (and $2\times10^{11}\ \mathrm{cm^{-2}}$ fluence), prior to gate-electrode patterning. The implantation was followed by rapid thermal annealing of the device at 1000$^{\circ}$C for five seconds. Although the device featured an on-chip microwave antenna, it was not utilised in this work. The device was wire bonded to a proprietary printed circuit board (PCB) and was stored within a copper sample enclosure. 
        The sample enclosure was placed within a Halbach array of neodymium magnets \cite{Adambukulam2021}, resulting in a magnetic field of approximately 1.1 T applied to the sample.
        The sample enclosure was securely attached to the mixing chamber plate of a Bluefors BF-LD400 dilution refrigerator, capable of reaching a base temperature of 20 mK. 

        A Stanford Research Systems SIM900 Mainframe containing SIM928 isolated DC-voltage sources supplied voltages to the SET gate electrodes, $V_{\mathrm{TG}}$, $V_{\mathrm{LB}}$, and $V_{\mathrm{RB}}$ via proprietary factor 8 resistive voltage dividers and to the source, $V_{\mathrm{bias}}$, via a factor 1000 voltage divider.

        The DC-voltages applied to the donor gates and plunger gate electrodes, $V_{\mathrm{DBR}}$, $V_{\mathrm{DBL}}$, $V_{\mathrm{DFR}}$, $V_{\mathrm{DFL}}$ and $V_{\mathrm{PL}}$, were controlled by National Instruments PXIe 4322 within a PXIe-1088. Additionally $V_{\mathrm{DFL}}$ and $V_{\mathrm{PL}}$ were connected to a Keysight M3300A arbitrary waveform generator (AWG), which was used to provide dynamic voltage signals, enabling fast acquisition of donor-SET charge stability diagrams. The AC and DC signals were combined at room temperature using impedance-matched combiners with a voltage division factor of 2.5 before being supplied to $V_{\mathrm{DFL}}$ and $V_{\mathrm{PL}}$ individually. DC-signals supplied to $V_{\mathrm{DBR}}$, $V_{\mathrm{DBL}}$, $V_{\mathrm{DFR}}$, and $V_{\mathrm{DFL}}$ were divided by a factor of 8. 
        
        The SET current was converted to a voltage using a FEMTO DLPCA-200 trans-impedance amplifier with a gain of 10$^7$ V/A and 50 kHz bandwidth. The signal was then passed through a Stanford Research Systems SIM910 JFET preamplifier, the gain was set to 1V/V as it acted as a ground connection breaker between the SET and what followed the preamplifier. The preamplifier was followed by a Stanford Research Systems SIM965 analogue 50 kHz low-pass band filter, and the converted signal was digitised and recorded by the Keysight M3300A at a sampling rate of 500 kHz. \texttt{donorsearch} features an open and flexible interface back-end that interfaces with SilQ \cite{Asaad2017} software for instrument control, which wraps around the QcoDeS \cite{Nielsen2019} acquisition framework and instrument drivers. 

        \subsection*{\texttt{donorsearch}'s workflow and parameters}

        \subsubsection*{Coarse tuning}\label{sup:Coarse}
        The algorithm checks the SET for turn-on by acquiring a current trace while simultaneously sweeping $V_{\mathrm{TG}}$, $V_{\mathrm{LB}}$ and $V_{\mathrm{RB}}$ from 0 V to 2 V. All other gate voltages are kept at 0 V. \texttt{donorsearch} checks for a factor of 100 change in current and a current range greater than 1 nA in the output trace. These are typical values for silicon SETs, as lower values usually indicate a faulty device. Therefore, if these conditions are not satisfied, turn-on is not achieved; the user is notified, and the algorithm stops. If successful, the algorithm extracts the approximate turn-on voltage by finding the lowest voltage corresponding to the current value 100 times higher than the minimum current (offset) value in the output trace. 
    
        Next, the algorithm tunes the SET by acquiring pinch-off current traces and applying the corresponding pinch-off voltages to the barrier gates. While $V_{\mathrm{TG}}$ is held constant at 2 V, the algorithm sequentially sweeps $V_{\mathrm{LB}}$ and $V_{\mathrm{RB}}$ from their upper to lower bounds and back, before proceeding to sweep the next gate.
        The SET current from each trace is first normalised, then smoothed with a Gaussian filter~\cite{Virtanen2020} and then fit to
        \begin{equation}
            f(x, A, B, C) = A(1 + \tanh(Bx + C))
        \end{equation}
        where $A$, $B$ and $C$ define the amplitude, slope, and shift offset respectively \cite{Darulova2020}. 
        
        This enables the extraction of the corresponding pinch-off voltages, which we define as the voltage at which the maximum of the second derivative of $f(x, A, B, C)$ occurs\cite{Botzem2018, Darulova2020}. If the curve fitting procedure is unsuccessful, \texttt{donorsearch} assumes that pinch-off is unattainable, notifies the user, and the algorithm stops. Otherwise, the algorithm tunes the SET to the extracted pinch-off voltages, while keeping $V_{\mathrm{TG}}$ at 2 V.

        \subsubsection*{Charge transition locator}\label{sup:locator}

        \begin{figure}[t] 
            \includegraphics[width=\columnwidth]{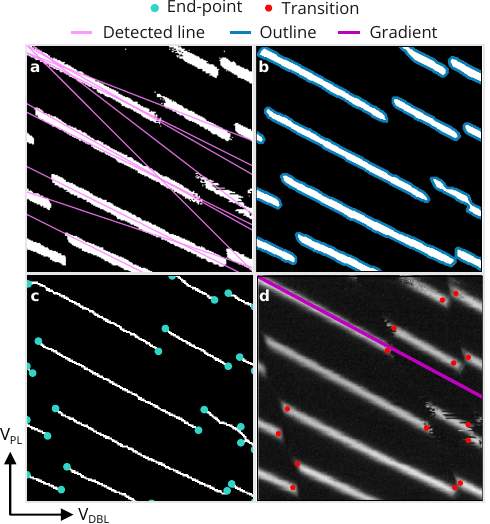}
            \caption{\label{fig:fig3_charge_transition_locator}Overview of the charge transition locator module, a computer vision and image analysis-based algorithm that identifies the charge transition locations in a charge stability diagram. (a) $I_{\mathrm{SET}}$ current heat maps are first threshold to a binarised image, and lines are detected using the Hough transform.  (b) After applying a Gaussian filter to the binarised image, the algorithm outlines the contours of the Coulomb peaks (blue solid lines).  (c) The medial axis and the end-points (turquoise) of these outlined contours are identified and provide a first estimation for the location of the donor-SET charge transitions. (d) The original input charge stability diagram is used as a mask to filter out false Coulomb peak end-points at the edges of the image and false end-points generated in the previous step (c). The filtered end-point locations (red) and extracted Coulomb peak gradients are returned and passed to the telegraphic signal classifier module.} 
        \end{figure}

        The charge transition locator module identifies the locations of the donor-SET charge transitions (see Fig. \ref{fig:fig3_charge_transition_locator}), which are later passed to the telegraphic signal classifier module and the fine tuning stage. The donor-SET charge stability diagrams acquired in the coarse tuning stage are first thresholded and binarised using Otsu's method \cite{Otsu1979}, using a threshold which minimises intra-class intensity variance. High current regions are assigned a pixel value of 1, and low current regions a pixel value of 0. The lines in the resulting binary image are extracted using a Hough transform \cite{Duda1972, Hough1962}, and the average gradient of the lines is inferred to be the slope of the SET-Coulomb peak (Fig. \ref{fig:fig3_charge_transition_locator}a, d). The slope value can then be used to construct a virtual gate for the calibration of spin readout through a three-level pulse sequence \cite{Morello2009, Morello2010}. 
        
        This binary image is then denoised using a Gaussian filter to smooth the edges of the Coulomb peaks. This avoids the algorithm being misguided by transitions displaying tunnelling effects (Fig. \ref{fig:fig3_charge_transition_locator}b). The contour outline of the Coulomb peaks is then extracted using a Marching Squares algorithm, the two-dimensional version of the Marching Cubes algorithm \cite{VanDerWalt2014, Lorensen1987} (solid blue lines in Fig.~\ref{fig:fig3_charge_transition_locator}b). 
        
        From these outlines, the medial axis of each Coulomb peak is calculated using the medial axis transform \cite{Blum1967},  resulting in a skeletonised image (Fig. \ref{fig:fig3_charge_transition_locator}c). The end-points of the medial axes, which are the candidate locations for the charge transition sites, are found by identifying pixels in the skeletonised binary image where the sum of nearest neighbour pixel values is equal to 1. These end-point candidates are then compared against the original binary image (e.g. {Fig.~\ref{fig:fig3_charge_transition_locator}a}), which serves as a mask to eliminate the end-points that may not correspond to real donor-SET charge transitions. We require that for each candidate end-point, the pixels located $\pm \Delta$ pixels along the average Coulomb peak slope must lie within the bounds of the binary image and must have opposite binary values, indicating a transition boundary. Moreover, no end-point can be within $r$ pixels of another. Note that in our charge stability diagrams, 1 pixel corresponds to a $2 \times 2$ mV area, and we set $\Delta$ and $r$ to 5 and 3 pixels, respectively. 
        
        Examples of end-points that are typically eliminated via this mask process are those near the edge of the image, or those arising from noise and tunnelling features that survived the previous filtering processes. Although this filtering approach results in some false negative charge transitions, we are willing to trade type I errors (false positives) for type II errors (false negatives) to maximise tuning speed. Finally, the estimated locations of the donor-SET charge transitions (red points in Fig. \ref{fig:fig3_charge_transition_locator}d) are passed to the telegraphic signal classifier.

        \subsubsection*{Telegraphic signal classifier}\label{sup:classifier}
            \begin{figure}[t] 
                \includegraphics[width=0.99\columnwidth]{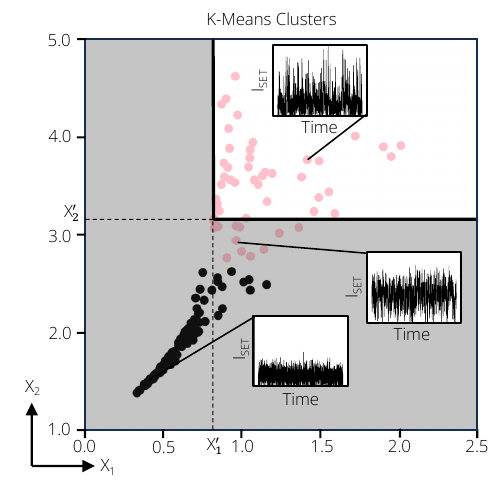}
                \caption{\label{fig:fig4_noise_classifier} K-means cluster map of the trace features, $X_1$ and $X_2$. The clusters do not satisfy the K-means assumption of circular groupings, but their rough identification can be used to separate regions of the feature space with relatively high accuracy. Current traces containing telegraphic signals are identified by the pink coloured point cluster, whereas traces containing mainly noise (identified by black points) are clustered in the bottom left of the map. The feature values used as classification thresholds by the classifier in the fine tuning stage ($X_{1}^{'}$, $X_{2}^{'}$), are found by locating the first pink point along the $X_1$ axis. Although this boundary may cause false negatives (middle current trace cutout), this method is fast and reduces the chances of false positives. Current traces acquired during the fine tuning stage that fall in the non-shaded region of the cluster map are then scored with the dynamic threshold method. Whereas traces which fall in the shaded region are considered as noise and assigned a score of 1.}
            \end{figure}

        The telegraphic signal classifier is built using unsupervised embedded learning (Fig. \ref{fig:fig4_noise_classifier}). It enables the algorithm to identify current traces containing telegraphic signals within the readout bandwidth. A 10-step random walk in gate-voltage space is initiated at each identified charge transition site using a Gaussian dice roll, e.g. the voltage change applied to each gate is sampled from a standard normal distribution with a standard deviation of 0.5 mV. These samples are added to the previous voltage value for all gates except $V_{\mathrm{TG}}$, which remains fixed at 2 V during the random walk. At each step of the random walk, a 30 ms current trace is measured with an integration time of 2 $\mu$s per point. All traces are then normalised by the minimum current observed across the entire set of measurements. From each trace, two features, $X_1$ and $X_2$, are extracted and used as inputs of a K-means clustering algorithm configured to identify two clusters. These features are defined as:
        \begin{equation}
            X_1 = \frac{m - n}{i}
        \end{equation}
        \begin{equation}
            X_2 = \frac{m}{n}
        \end{equation}
        
        where $m$, $n$, and $i$ are respectively the maximum, minimum, and mean of the current trace. Before estimating $X_1$ and $X_2$, we downsample the current trace to an effective integration time of 50 $\mu$s per point, to improve the signal-to-noise ratio.

        To illustrate the rationale behind the clustering algorithm, Fig.~\ref{fig:fig4_noise_classifier} shows current traces from a representative run (each represented as a point), plotted according to their extracted features $X_1$ and $X_2$. Current traces without telegraphic signals (shown as black points) cluster in the bottom left of the K-means map. This is because, when no electron tunnelling events are detected, the small difference between $m$ and $n$ results in near-zero values for $X_1$, while $X_2$ tends towards 1. On the other hand, current traces with telegraphic signals (shown as pink points) tend to spread toward the upper right quadrant of the feature map. This is because as the difference between $m$ and $n$ increases due to tunnelling events being detected, $X_2$ rises while $X_1$ tend to spread because the average current value in the denominator can vary due to different relative occupations of the two levels.
        Since the clusters do not satisfy the standard K-means assumptions \cite{Bock1996} of circular clusters, we use what is provided by K-means to separate the feature space into two different regions: one containing telegraphic signals and one containing noise.

        To extract the boundary that separates the two regions efficiently, the algorithm scans along the $X_1$ axis and takes the coordinates of the first pink cluster point found ($X_{1}^{'}$, $X_{2}^{'}$) as the threshold features. The rejection area, shaded in grey, although not defined by the true separation boundary between the clusters, has the benefit of reducing the number of false positive current traces, since traces with larger $X_1$ are those where the average current ($i$) is closer to the minimum current ($n$). 

        During the search for telegraph signals in the fine tuning stage, $X_1$ and $X_2$ are first calculated and compared to $X_{1}^{'}$ and $X_{2}^{'}$ for each current trace acquired. If both $X_1$ and $X_2$ are greater than $X_{1}^{'}$ and $X_{2}^{'}$ respectively, the current trace is considered to contain telegraphic signals; otherwise, the acquired current trace is considered as noise and rejected. Our results indicate that classification via direct feature comparison is both reliable and significantly faster than reapplying the clustering algorithm to new data, since its runtime scales with the number of samples.

        \texttt{donorsearch} uses a Gaussian process Bayesian optimisation as a default search method during the fine tuning stage. Random search is a second search method (available within the current code), which we used as a benchmark against the machine learning search-based method. Random search is carried out by performing uniform random samples within the bounds of the gate voltages. The Gaussian process Bayesian optimiser \cite{Head2018} aims to minimise the score. It uses the Matérn 5/2 kernel and relies on negative Expected Improvement \cite{Jones1998} as its acquisition function, which aims to minimise over the posterior distribution. The kernel's characteristic length scales ${\ell}_{q}$, where $q = 1, \ldots, N$ where $N = 7$ corresponding to the 7 gates controlled by the algorithm, and ${\sigma}_f$ the covariance amplitude, are optimised using maximum likelihood estimation.
        We chose negative Expected Improvement as the desired acquisition function because the charge transition locator module reduces the overall search space for the fine tuning stage. As a result, less explorative sampling techniques are required, such as the hypersurface sampling technique used in Refs. \cite{Severin2021, Lennon2019, VanStraaten2022}.
        The optimiser acquires the first ten samples randomly before approximating the voltage space with a Gaussian process.

        During the fine tuning stage, each current trace acquired is scored for telegraphic signals. The scoring method relies on applying a double Gaussian fit to the current trace and extracting the split between the two Gaussians as the current threshold. The software used to apply the double Gaussian fit is the Quantum Technology Toolbox (QTT) developed by QuTech \cite{Eendebak2018}, which relies on the Non-Linear Least Square Minimisation and Curve-Fitting for Python (LMFIT) package \cite{Newville2023}. The separation, $d$, of the two Gaussians is defined as \cite{Eendebak2018},

        \begin{equation}
            d = \frac{\mu_2 - \mu_1}{(\sigma_1 + \sigma_2)}
        \end{equation}

        and the split, $s$, is

        \begin{equation}
            s = \mu_1 + d \sigma_1
        \end{equation}

        where $\mu_{1/2}$ and $\sigma_{1/2}$ are the mean and standard deviation of the respective Gaussians and $\mu_2 > \mu_1$.
    
        \subsubsection*{Labelling procedure}\label{sup:labels}
        To check the algorithm's performance, two humans labelled the current traces claimed by the algorithm to contain telegraphic signals. Current traces from all the experimental runs were gathered and shuffled randomly, then labelled by each human separately. This was done to reduce any bias towards the different branches of \texttt{donorsearch}'s fine tuning stage (machine-learning enabled or random search) and peer bias from other human labellers. Fine tuning times were calculated using a Bayesian model of multi-labeller statistics as in Ref. \cite{Moon2020}.        

        \begin{figure}[ht!]
            \includegraphics[width=0.99\columnwidth]{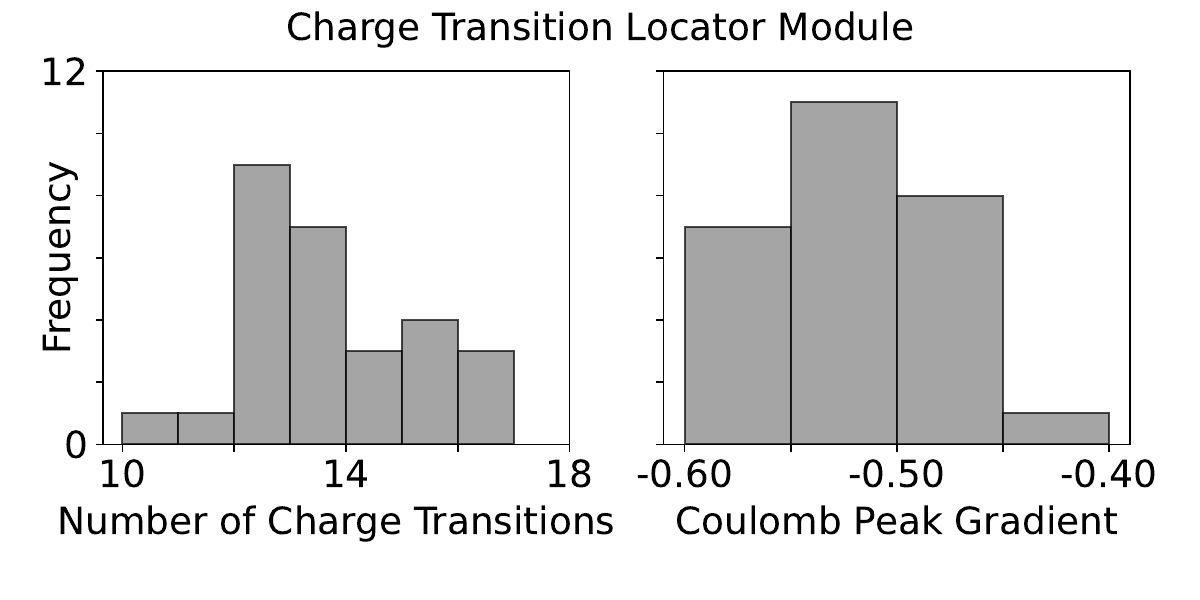}
            \caption{\label{fig:Suppfig1_transition_locator_stats}
            Charge transition locator module performance. Bar charts of the number of charge transitions and range of Coulomb peak gradients extracted by \texttt{donorsearch} across all experimental runs.} 
        \end{figure}

        \begin{figure}[ht!]
            \includegraphics[width=0.99\columnwidth]{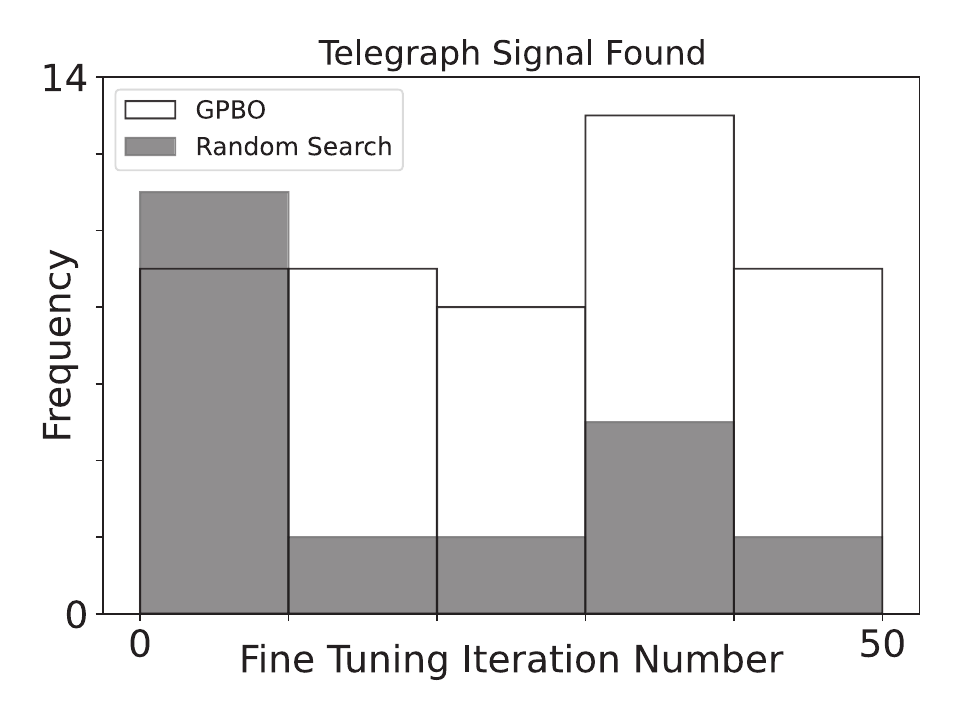}
            \caption{\label{fig:Suppfig2_fine_tuning_teration_success}
            Search for the telegraph signal success rate. Bar chart showing iteration and the respective search methods in the fine tuning stage, Gaussian process Bayesian optimisation (GPBO) and random search, finds telegraph signals based on the unanimous human labels. }
        \end{figure}

        \subsection*{Supplementary Tables}\label{sup:tables}

    \begin{table*}[ht!]
        \caption{\label{tab:donorsearch_labels} Total number of current traces labelled as positive (i.e. containing random telegraph signal) found by each labeller (Labeller 1 and 2) for all experimental runs, including Gaussian process Bayesian optimisation (GPBO) and random search implementations for the fine tuning stage, alongside a breakdown of the experimental time for each stage of the algorithm. The run marked with an asterisk (*) was used as a sample run to gauge the performance of the noise classifier.
        } 
        \begin{ruledtabular}

        \begin{tabular}{ccccccc}
        Experiment           & Coarse (s) & Handshake (s) & Fine (s) & Total (s) & Labeller 1 & Labeller 2 \\
        \hline
        GPBO run 1           & 333.32     & 147.12        & 623.07   & 1,103.51  & 1          & 1          \\
        GPBO run 2           & 316.41     & 142.93        & 582.61   & 1,041.95  & 6          & 6          \\
        GPBO run 3           & 318.11     & 143.53        & 619.69   & 1,081.34  & 4          & 4          \\
        GPBO run 4           & 317.88     & 115.84        & 415.53   & 849.25    & 5          & 5          \\
        GPBO run 5           & 315.12     & 128.10        & 490.22   & 933.44    & 6          & 5          \\
        GPBO run 6           & 315.98     & 112.02        & 371.92   & 799.92    & 5          & 3          \\
        GPBO run 7           & 315.09     & 104.42        & 443.48   & 862.99    & 4          & 3          \\
        GPBO run 8           & 316.42     & 129.46        & 556.69   & 1,002.57  & 3          & 3          \\
        GPBO run 9           & 316.06     & 112.79        & 417.17   & 846.01    & 4          & 3          \\
        GPBO run 10          & 316.96     & 112.83        & 474.93   & 904.72    & 4          & 3          \\
        GPBO run 11          & 323.02     & 118.63        & 502.52   & 944.16    & 1          & 1          \\
        GPBO run 12          & 321.69     & 110.41        & 376.04   & 808.15    & 6          & 6          \\
        GPBO run 13          & 322.10     & 124.85        & 547.13   & 994.08    & 3          & 3          \\
        GPBO run 14          & 324.42     & 107.95        & 432.62   & 864.99    & 4          & 2          \\
        \hline
        Random Search run 1  & 317.63     & 134.81        & 497.57   & 950.00    & 3          & 1          \\
        Random Search run 2  & 315.68     & 133.22        & 496.32   & 945.22    & 1          & 1          \\
        Random Search run 3  & 314.68     & 102.98        & 399.42   & 817.07    & 1          & 1          \\
        Random Search run 4  & 314.46     & 104.05        & 370.24   & 788.75    & 2          & 2          \\
        Random Search run 5  & 317.86     & 119.31        & 421.48   & 858.65    & 2          & 2          \\
        Random Search run 6  & 317.23     & 97.81         & 304.15   & 719.19    & 3          & 2          \\
        Random Search run 7  & 317.26     & 107.66        & 360.96   & 785.89    & 3          & 2          \\
        Random Search run 8  & 316.50     & 108.78        & 422.88   & 848.17    & 2          & 1          \\
        Random Search run 9  & 317.68     & 156.31        & 506.86   & 980.85    & 3          & 1          \\
        Random Search run 10 & 317.59     & 113.57        & 349.36   & 780.52    & 4          & 3          \\
        Random Search run 11 & 319.89     & 147.13        & 583.70   & 1,050.73  & 1          & 0          \\
        Random Search run 12 & 320.22     & 109.64        & 422.78   & 852.64    & 2          & 2          \\
        Random Search run 13 & 321.80     & 119.52        & 483.90   & 925.23    & 1          & 1          \\
        Random Search run 14 & 322.66     & 131.63        & 430.16   & 884.45    & 3          & 3         
        \end{tabular}
        \end{ruledtabular}
    \end{table*}

\begin{table*}[ht!]
{\color{black}
\begin{ruledtabular}
\caption{\label{tab:noise_classifier_confusion_matrixes} \textcolor{black}{Confusion matrices of the noise classifier tested from one of the experimental runs when using unanimous human labels to gauge its performance as a noise classifier and as a telegraphic signal classifier.}}

\begin{tabular}{lccllcc}
                              & \multicolumn{2}{c}{Noise Classifier} &  &                               & \multicolumn{2}{c}{Telegraphic Signal Classifier} \\
                              & Pred. Neg.        & Pred. Pos.       &  &                               & Pred. Neg.             & Pred. Pos.             \\
\multicolumn{1}{c}{True Neg.} & 39                & 15               &  & \multicolumn{1}{c}{True Neg.} & 111                    & 4                      \\
\multicolumn{1}{c}{True Pos.} & 1                 & 105              &  & \multicolumn{1}{c}{True Pos.} & 9                      & 36                    
\end{tabular}
\end{ruledtabular}
}
\end{table*}

\begin{table*}[ht!]
{\color{black}
\begin{ruledtabular}
\caption{\label{tab:search_rts_confusion_matrixes} \textcolor{black}{Confusion matrices of the current traces deemed by \texttt{donorsearch} to contain random telegraph signal at the end of search for random telegraph signal relying on Gaussian process Bayesian optimisation (GPBO) or random search and overall experimental runs. Confusion matrices are shown in the unanimous form where all human labellers agree on the presence of random telegraph signal and in the single labeller form where a signal labeller confirmed the presence of random telegraph signal in a given current trace.}}

\begin{tabular}{lccllccllcc}
                              & \multicolumn{2}{c}{GPBO - Unanimous}        &  &                               & \multicolumn{2}{c}{Random Search - Unanimous} &  &                               & \multicolumn{2}{c}{Overall - Unanimous}     \\
                              & Pred. Neg.           & Pred. Pos.           &  &                               & Pred. Neg.            & Pred. Pos.            &  &                               & Pred. Neg.           & Pred. Pos.           \\
\multicolumn{1}{c}{True Neg.} & 0                    & 14                   &  & \multicolumn{1}{c}{True Neg.} & 0                     & 22                    &  & \multicolumn{1}{c}{True Neg.} & 0                    & 36                   \\
\multicolumn{1}{c}{True Pos.} & 0                    & 48                   &  & \multicolumn{1}{c}{True Pos.} & 0                     & 22                    &  & \multicolumn{1}{c}{True Pos.} & 0                    & 70                   \\
                              & \multicolumn{1}{l}{} & \multicolumn{1}{l}{} &  &                               & \multicolumn{1}{l}{}  & \multicolumn{1}{l}{}  &  &                               & \multicolumn{1}{l}{} & \multicolumn{1}{l}{} \\
                              & \multicolumn{2}{c}{GPBO - Single}           &  &                               & \multicolumn{2}{c}{Random Search - Single}    &  &                               & \multicolumn{2}{c}{Overall - Single}        \\
                              & Pred. Neg.           & Pred. Pos.           &  &                               & Pred. Neg.            & Pred. Pos.            &  &                               & Pred. Neg.           & Pred. Pos.           \\
\multicolumn{1}{c}{True Neg.} & 0                    & 6                    &  & \multicolumn{1}{c}{True Neg.} & 0                     & 13                    &  & \multicolumn{1}{c}{True Neg.} & 0                    & 19                   \\
\multicolumn{1}{c}{True Pos.} & 0                    & 56                   &  & \multicolumn{1}{c}{True Pos.} & 0                     & 31                    &  & \multicolumn{1}{c}{True Pos.} & 0                    & 87                  
\end{tabular}
\end{ruledtabular}
}
\end{table*}

\clearpage

\end{document}